\documentclass[10pt,a4letter]{article}
\usepackage[english]{babel}
\usepackage{comment}

\usepackage{amsmath}
\usepackage{amssymb}
\usepackage{dsfont}
\usepackage{graphicx,epsfig}
\usepackage{slashed}
\usepackage[mathscr]{eucal}
\usepackage{placeins}

\usepackage{color}
\usepackage[numbers,sort&compress]{natbib}
\usepackage{multirow}
\usepackage{tabularray}
\usepackage{tabularx}
\UseTblrLibrary{booktabs}
\usepackage{float}

\usepackage{slashed}
\usepackage{braket}

\usepackage{scalefnt,pstricks}

\usepackage{xspace}
\usepackage{setspace}
\usepackage{xstring}
\usepackage{mathrsfs}
\usepackage{amsbsy}
\usepackage{makecell}
\usepackage{textcomp}

\usepackage{lipsum}
\usepackage[colorlinks=true
,urlcolor=blue
,anchorcolor=blue
,citecolor=blue
,filecolor=blue
,linkcolor=blue
,menucolor=blue
,pagecolor=blue
,linktocpage=true
,pdfproducer=medialab
,pdfa=true
]{hyperref}
\usepackage[capitalise]{cleveref}
\usepackage{shuffle}

\newcommand\as{\alpha_{\mathrm{S}}}
\newcommand\muR{\ensuremath{\mu_{\rm R}}\xspace}
\newcommand\muF{\ensuremath{\mu_{\rm F}}\xspace}
\newcommand\muIR{\ensuremath{\mu_{\rm IR}}\xspace}
\newcommand\M{{\cal M}}
\newcommand\OpenLoops{{\sc OpenLoops}\xspace}
\newcommand\Recola{{\sc Recola}\xspace}
\newcommand\eps{\ensuremath{\epsilon}\xspace}
\newcommand\bareps{\ensuremath{\bar{\eps}}\xspace}
\newcommand\Matrix{{\sc Matrix}\xspace}
\newcommand\AMFlow{{\sc AMFlow}\xspace}
\newcommand\qqttH{\ensuremath{q \bar q\to t\bar t H}\xspace}
\newcommand\ttH{\ensuremath{t \bar t H}\xspace}
\newcommand\ttW{\ensuremath{t \bar t W}\xspace}
\newcommand\betatwo{\ensuremath{\beta^2}\xspace}
\newcommand\fracstt{\ensuremath{\operatorname{frac}_{s_{t\bar t}}}\xspace}
\newcommand\cthetaH{\ensuremath{\cos \theta_H}\xspace}
\newcommand\cthetat{\ensuremath{\cos \theta_t}\xspace}
\newcommand\phit{\ensuremath{\phi_t}\xspace}
\newcommand\Hone{\ensuremath{H^{(1)}}\xspace}
\newcommand\Hn{\ensuremath{H^{(n)}}\xspace}
\newcommand\HoneML{\ensuremath{H^{(1)}_{\mathrm{ML}}}\xspace}
\newcommand\HoneLCAML{\ensuremath{H^{(1)}_{\mathrm{LCA,ML}}}\xspace}
\newcommand\HtwoLCAML{\ensuremath{H^{(2)}_{\mathrm{LCA,ML}}}\xspace}
\newcommand\HnLCAML{\ensuremath{H^{(n)}_{\mathrm{LCA,ML}}}\xspace}
\newcommand\absHoneML{\ensuremath{|\HoneML|}\xspace}
\newcommand\absHoneLCAML{\ensuremath{|\HoneLCAML|}\xspace}
\newcommand\absHtwoLCAML{\ensuremath{|\HtwoLCAML|}\xspace}
\newcommand\absHnLCAML{\ensuremath{|\HnLCAML|}\xspace}
\newcommand\dsHone{\ensuremath{\mathrm{d}\sigma_{\hspace*{-0.2em}\Hone}}\xspace}
\newcommand\dsHoneML{\ensuremath{\mathrm{d}\sigma_{\hspace*{-0.2em}\HoneML}}\xspace}
\newcommand\dsHoneLCAML{\ensuremath{\mathrm{d}\sigma_{\hspace*{-0.2em}\HoneLCAML}}\xspace}
\newcommand\dsHtwoLCAML{\ensuremath{\mathrm{d}\sigma_{\hspace*{-0.2em}\HtwoLCAML}}\xspace}
\newcommand\dsHnLCAML{\ensuremath{\mathrm{d}\sigma_{\hspace*{-0.2em}\HnLCAML}}\xspace}
\newcommand\dsabsHoneML{\ensuremath{\mathrm{d}\sigma_{\hspace*{-0.1em}\absHoneML}}\xspace}
\newcommand\dsabsHoneLCAML{\ensuremath{\mathrm{d}\sigma_{\hspace*{-0.1em}\absHoneLCAML}}\xspace}
\newcommand\dsabsHtwoLCAML{\ensuremath{\mathrm{d}\sigma_{\hspace*{-0.1em}\absHtwoLCAML}}\xspace}
\newcommand\dsabsHnLCAML{\ensuremath{\mathrm{d}\sigma_{\hspace*{-0.1em}\absHnLCAML}}\xspace}
\newcommand\dsabsHoneLCAMLminusHoneML{\ensuremath{\mathrm{d}\sigma_{\hspace*{-0.1em}|\HoneLCAML-\HoneML|}}\xspace}
\newcommand\roneLCA{\ensuremath{r^{(1)}_{\mathrm{LCA}}}\xspace}
\newcommand\deltaoneHoneLCA{\ensuremath{\delta^{(1),H^{(1)}}_{\mathrm{LCA}}}\xspace}
\newcommand\deltatwoHoneLCA{\ensuremath{\delta^{(2),H^{(1)}}_{\mathrm{LCA}}}\xspace}
\newcommand\deltaonemuIRLCA{\ensuremath{\delta^{(1),\muIR}_{\mathrm{LCA}}}\xspace}
\newcommand\deltatwomuIRLCA{\ensuremath{\delta^{(2),\muIR}_{\mathrm{LCA}}}\xspace}
\newcommand\deltatwoLCA{\ensuremath{\delta^{(2)}_{\mathrm{LCA}}}\xspace}
\newcommand\HtwoLCA{\ensuremath{H^{(2)}_{\mathrm{LCA}}}\xspace}
\newcommand\HtwoSAMA{\ensuremath{H^{(2)}_{\mathrm{SA+MA}}}\xspace}
\newcommand\sHone{\ensuremath{\sigma_{\hspace*{-0.2em}\Hone}}\xspace}
\newcommand\sHoneLCAML{\ensuremath{\sigma_{\hspace*{-0.2em}\HoneLCAML}}\xspace}
\newcommand\sHtwoLCA{\ensuremath{\sigma_{\hspace*{-0.2em}\HtwoLCA}}\xspace}
\newcommand\sHtwoSAMA{\ensuremath{\sigma_{\hspace*{-0.2em}\HtwoSAMA}}\xspace}

\begin{document}
\begin{titlepage}
\begin{flushright}
ZU-TH 44/26 \\
\end{flushright}

\vspace*{0.5cm}

\begin{center}
{\Large \bf Two-loop QCD contribution to the quark-initiated production\\[0.3cm]of a top-antitop quark pair and a Higgs boson at the LHC}
\end{center}
\vskip 0.5cm
\par \vspace{2mm}
\begin{center}
{\bf Xiang~Chen}, {\bf Massimiliano Grazzini} and {\bf Stefan Kallweit} 

Physik-Institut, Universit\"at Z\"urich, Winterthurerstrasse 190, 8057 Z\"urich, Switzerland\\[0.25cm]
\end{center}

\vskip 0.5cm
\par \vspace{2mm}
\begin{center} {\large \bf Abstract}

\end{center}
\begin{quote}
\pretolerance 10000

We present the first direct computation of the two-loop QCD virtual corrections to the
production of a top--antitop quark pair in association with a Higgs boson (\ttH) at the LHC.
We focus on the quark--antiquark channel and perform the calculation in the generalised leading-colour approximation.
We combine state-of-the-art numerical methods for multiloop calculations with machine learning techniques to
construct a surrogate amplitude suitable for accurate cross section computations.
We compare our numerical results with those previously obtained using dynamical approximations of the two-loop amplitude,
finding agreement within their respective uncertainties.

\end{quote}

\vspace*{\fill}
October 2026
\end{titlepage}

\section{Introduction}
The associated production of a Higgs boson with a top–antitop quark pair (\ttH) was observed by the
ATLAS and CMS experiments at the LHC in 2018 \cite{ATLAS:2018mme,CMS:2018uxb}.
Among the various Higgs boson production channels, this process is particularly relevant as a direct probe of the top-quark Yukawa coupling.
Precise measurements of this coupling require accurate theoretical predictions to which the data can be compared.
Consequently, \ttH production has been the subject of extensive theoretical work over recent decades.

The first theoretical studies of \ttH production date back more than forty years~\cite{Ng:1983jm,Kunszt:1984ri},
while next-to-leading-order (NLO) QCD corrections have been computed in
Refs.~\cite{Beenakker:2001rj,Beenakker:2002nc,Reina:2001sf,Reina:2001bc,Dawson:2002tg,Dawson:2003zu}.
These calculations were matched to parton showers in Refs.~\cite{Frederix:2011zi,Garzelli:2011vp,Hartanto:2015uka}.
The NLO electroweak (EW) corrections were first calculated in Ref.~\cite{Frixione:2014qaa} and subsequently combined with NLO QCD
contributions, within the narrow-width approximation for the top-quark decays, in Refs.~\cite{Yu:2014cka,Frixione:2015zaa}.

The resummation of soft-gluon contributions close to the partonic threshold was addressed in
Refs.~\cite{Kulesza:2015vda,Broggio:2015lya,Broggio:2016lfj,Kulesza:2017ukk},
while soft and Coulomb corrections were considered in Ref.~\cite{Ju:2019lwp}.
The resummed results up to next-to-next-to-leading logarithmic (NNLL) accuracy, matched to the fixed-order NLO results,
were further improved by including EW corrections~\cite{Broggio:2019ewu,Kulesza:2020nfh}.
The NLO QCD corrections to $\ttH$ production with off-shell top quarks and leptonic $W$-boson decays
were studied in Refs.~\cite{Denner:2015yca, Stremmer:2021bnk}, while full off-shell effects
including the \mbox{$H \to b \bar{b}$} decay were investigated in Refs.~\cite{Denner:2020orv,Bevilacqua:2022twl}.
A combination of the NLO QCD and EW effects was presented in Ref.~\cite{Denner:2016wet}.

Theoretical predictions based on NLO QCD are affected by an uncertainty of
about $10\%$~\cite{LHCHiggsCrossSectionWorkingGroup:2016ypw}.
Including next-to-next-to-leading order (NNLO) QCD corrections is therefore essential to match the experimental
precision expected by the end of the HL-LHC phase.
The bottleneck for achieving NNLO accuracy is represented by the two-loop contribution. 
The \mbox{$2\to 3$} kinematics, with different masses for both external and internal legs, make this calculation
one of the most demanding to date for modern amplitude techniques.
The difficulty is reflected in the closely related $\ttW$ process, for which progress has been achieved
only very recently~\cite{Becchetti:2026awn,Becchetti:2026yxl}.
The $\ttH$ process presents an additional challenge: the coupling of the Higgs boson to the top quark gives rise to
further internal massive propagators, making the two-loop calculation even more complex.

First NNLO predictions for $\ttH$ production have been presented in Refs.~\cite{Catani:2022mfv,Devoto:2024nhl}.
These calculations relied on approximations of the two-loop contribution, in which the finite remainder was estimated
either in the soft-Higgs or in the high-energy limit.%
\footnote{Other approximations considered in the literature
either neglect power-suppressed contributions in the Higgs boson mass~\cite{Wang:2024pmv},
or treat the Higgs boson as a collinear parton radiated off a top quark.
The latter approach was used in early NLO calculations \cite{Dawson:1997im} of \ttH production,
and more recently in Refs.~\cite{Brancaccio:2021gcz,Brancaccio:2026rmo}.}
Combined with soft-gluon resummation and EW corrections, these predictions
provide the current theory reference for this process \cite{Balsach:2025jcw}.
The approximations of Refs.~\cite{Catani:2022mfv,Devoto:2024nhl}
were also used to match the NNLO calculation to a parton shower~\cite{Biello:2026nhj}.

Although the uncertainties associated with such dynamical approximations appear to be under control,
a direct calculation of the two-loop amplitude remains essential to validate the reliability of this approach
and ultimately eliminate the ensuing systematic uncertainties.

Work in this direction has been pursued by several groups.
The one-loop corrections to the \mbox{$gg\to t{\bar t}H$} amplitude have been computed up to ${\cal O}(\epsilon^2)$~\cite{Buccioni:2023okz}.
The master integrals~(MIs) involving closed light-fermion loops were evaluated in Ref.~\cite{FebresCordero:2023pww}.
First numerical results for the quark-initiated \ttH two-loop amplitudes containing all closed fermion loops
were presented in Ref.~\cite{Agarwal:2024jyq}.

In this Letter we make a decisive step towards a direct computation of $\ttH$ amplitudes at two-loop order
by reporting the first calculation of the \qqttH amplitude in the generalised leading-colour approximation~(LCA).
We also present the quantitative impact of the computed corrections and
compare them with those obtained in Ref.~\cite{Devoto:2024nhl}.

\section{Calculation}

The results we are going to present refer to purely virtual contributions to the hadronic cross section,
and are obtained within the \Matrix framework~\cite{Grazzini:2017mhc}.
Exact tree-level and one-loop amplitudes are computed with
\OpenLoops~\cite{Cascioli:2011va, Buccioni:2017yxi,Buccioni:2019sur} and
\Recola~\cite{Actis:2016mpe,Denner:2017wsf,Denner:2016kdg}.
The finite part of the $n$-loop \qqttH virtual amplitude contributes to the N$^n$LO correction in the form
\begin{equation}
\label{eq:sigmaH}
\mathrm{d}\sigma_{H^{(n)}}=\left(\frac{\as(\muR)}{4\pi}\right)^n H^{(n)}\times \mathrm{d}\sigma_{\rm LO} \,,
\end{equation}
where $\as(\muR)$ is the QCD coupling at the renormalisation scale $\muR$, $\mathrm{d}\sigma_{\rm LO}$ is the
Born-level cross section, and the hard-virtual coefficient $H^{(n)}$ is defined as
\begin{equation}
\label{eq:Hn}
H^{(n)}= \frac{2 \, {\rm Re}\left[\M^{(n)}_{\rm fin}\, \M^{(0)*} \right]}{\left|\M^{(0)}\right|^2}\,.
\end{equation}
In Eq.~(\ref{eq:Hn}), $\M^{(n)}_{\rm fin}$ is the renormalised scattering amplitude for the \qqttH process,
and a sum over polarisations is understood.
The superscript `$(n)$' denotes the $n$-th perturbative order ($n=1,2$ here) in an expansion in powers of $\as(\muR)/(4\pi)$,
while the subscript `${\rm fin}$' indicates that infrared (IR) singularities have been subtracted according
to the prescription of Ref.~\cite{Ferroglia:2009ii} at the scale $Q$, with $Q$ the invariant mass of the \ttH system.
The calculation of the two-loop amplitude $\M_{\rm fin}^{(2)}$ is the crucial element of this work,
and its technical details will be reported elsewhere~\cite{inprep}.
Here, we limit ourselves to summarising the main points.

We consider the partonic process
\begin{equation}
\label{eq:process}
q(p_1)+\bar{q}(p_2)\to t(p_3)+\bar{t}(p_4)+H(p_5) \,,
\end{equation}
and work in the LCA%
\footnote{Note that the LCA is applied only to $H^{(n)}$, while $\mathrm{d}\sigma_{\rm LO}$ is computed exactly.}%
, i.e.\ we keep only the terms of the two-loop amplitude proportional to
$N_c^2$, $N_c n_l$ and $n_l^2$, where \mbox{$N_c=3$} is the number of colours and \mbox{$n_l=5$} the
number of light quarks circulating in the loops.
The Feynman diagrams are generated with \textsc{FeynArts}~\cite{Hahn:2000kx} and \textsc{Qgraf}~\cite{Nogueira:1991ex}.
The interference between the two-loop and tree-level amplitudes is
computed using a private \textsc{Mathematica} package that performs
the contraction of both colour and Lorentz indices and the Dirac algebra.

To reduce the complexity of the integral reduction, the pole mass of the top quark $m_t$
and the Higgs boson mass $m_H$ are fixed to rationalised numerical values, normalised to the top-quark mass,
i.e.\ \mbox{$m_t^2=1$} and \mbox{$m_H^2/m_t^2=661/1257$}, which approximately corresponds to \mbox{$m_t=172.5$} GeV
and \mbox{$m_H=125.09$} GeV.
In addition, throughout the entire calculation, the dimensional regulator \eps (\mbox{$\eps=(4-d)/2$},
with $d$ the number of space-time dimensions) is assigned the numerical values $\pm \bareps$,
with \mbox{$\bareps=10^{-3}$}~\cite{Bi:2023bnq,Bi:2025oga,Li:2025bsq,Becchetti:2026awn,Becchetti:2026yxl}.
As a consequence, only five independent Mandelstam invariants
are kept symbolic in the generation of the reduction identities.
This drastically reduces the algebraic complexity compared to a fully symbolic reduction,
at the expense of evaluating the expressions numerically at two values of \eps.

The reduction of the two-loop amplitude to MIs is based on linear relations derived from
integration-by-parts (IBP) identities~\cite{Tkachov:1981wb,Chetyrkin:1981qh,Laporta:2000dsw}. We consider two classes of integrals, distinguished by their origin:
those appearing in the amplitudes, which are typically of higher rank, and those entering the differential equations (DEs)~\cite{Kotikov:1990kg,Remiddi:1997ny}
for the MIs, which usually have higher powers of the propagators. To construct the systems used in the reductions, we use two packages,
\textsc{NeatIBP}~\cite{Wu:2023upw,Wu:2025aeg} and \textsc{Blade}~\cite{Guan:2024byi}.
\textsc{NeatIBP} generates compact IBP systems, while \textsc{Blade} searches for a
\textit{block-triangular form}~\cite{Guan:2019bcx} that serves as an alternative, substantially smaller linear system for the reduction, 
with five independent Lorentz invariants remaining symbolic.
Their performance depends on both the structure of the integrals and the integral family, so we compare the two packages for each class--family combination and adopt whichever performs the reduction faster.

Once all the systems have been obtained,
we use \textsc{Ratracer}~\cite{Magerya:2022hvj}, which relies on \textsc{FireFly}~\cite{Klappert:2019emp,Klappert:2020aqs},
to reconstruct the rational reduction coefficients from
finite-field evaluations~\cite{vonManteuffel:2014ixa,Peraro:2016wsq}.
The two classes require different reconstruction strategies. For integrals appearing in the amplitudes,
we reconstruct the coefficients expressing these integrals in terms of MIs separately at
each phase-space point~\cite{Peraro:2019okx}.
For integrals entering the DEs, we parameterise the path from the boundary point to the target point
and reconstruct the coefficients as rational functions of the path parameter.
Since the MIs are obtained by solving the DEs along this path,
this one-dimensional reconstruction is sufficient.

The MIs are then evaluated by numerically solving the DEs
with the \textsc{Mathematica} implementation of \AMFlow's DE solver~\cite{Liu:2017jxz,Liu:2021wks,Liu:2022chg,Huang:2026rjb},
using boundary conditions computed with \AMFlow to 60-digit precision.
When setting up these equations, the path is required to lie within the physical region to avoid crossing any physical singularities.
The only singularities encountered along the path are spurious poles, whose number increases because of the choice of
MIs according to the Laporta algorithm~\cite{Laporta:2000dsw}.
A similar approach has been successfully applied in the recent calculation of the two-loop amplitude
for \ttW production~\cite{Becchetti:2026awn,Becchetti:2026yxl}.
After performing renormalisation and IR subtraction, the $H^{(n)}$
evaluations at \mbox{$\eps = \pm \bareps$} are averaged
to obtain the hard-virtual coefficient at \mbox{$\eps=0$}.
The \mbox{$\mathcal{O}(\bareps^2)$} truncation error, estimated by squaring the relative \mbox{$\mathcal{O}(\bareps)$} difference, \mbox{$\delta=|(H^{(n)}_{+}-H^{(n)}_{-})/(H^{(n)}_{+}+H^{(n)}_{-})|^2$}, never exceeds $3\times10^{-4}$.

Next, we repeat the computation for a selection of phase-space points and
extend the results to the entire phase space.
A comprehensive discussion of the main methods for amplitude interpolation can be found in Ref.~\cite{Breso-Pla:2024pda},
where the authors find that \textit{machine learning} (ML) is the most practical choice for multi-dimensional cases,
such as \mbox{$2\to3$} processes, at least if the precision requirement is not extremely high.
We therefore train a surrogate amplitude~\cite{Aylett-Bullock:2021hmo,Maitre:2021uaa,Badger:2022hwf,Maitre:2023dqz,Spinner:2024hjm,Brehmer:2024yqw,Breso-Pla:2024pda,Bahl:2024gyt,Bahl:2025xvx,Villadamigo:2025our,Beccatini:2025tpk,Bahl:2026qaf}
using a \textit{multi-layer perceptron}~(MLP)~\cite{rosenblatt1958perceptron,rumelhart1986learning},
with the loss function defined as
\mbox{$\mathcal{L}=\int|\tilde{f}(\vec{x})-f(\vec{x})|\,\mathrm{d}\vec{x}\big/\int|f(\vec{x})|\,\mathrm{d}\vec{x}$}, where \mbox{$f(\vec{x})=a(\vec{x})w(\vec{x})$}, with $a(\vec{x})$ ($a_{\rm LO}(\vec{x})$) the numerator (denominator) in Eq.~(\ref{eq:Hn}) and
$w(\vec{x})$ the weight function, which includes both the appropriate phase-space measure and the parton densities~\cite{Breso-Pla:2024pda}.
In Monte Carlo integration, the loss can be expressed as
\begin{equation}
\label{eq:lossfunction}
\mathcal{L}=\dfrac{\sum_{i=1}^{m}|(\tilde{a}_i-a_i)/a_{{\rm LO},i}|}{\sum_{i=1}^{m}|a_i/a_{{\rm LO},i}|}=\dfrac{\sum_{i=1}^{m}|\tilde{H}_i^{(n)}-H_i^{(n)}|}{\sum_{i=1}^{m}|H_i^{(n)}|}\,,
\end{equation}
where the sums run over the $m$ points of a dataset, sampled according to $a_{\rm LO}w$, and a tilde denotes a surrogate quantity.
We generate three such datasets: {\it training}, {\it validation} and {\it test}.
The validation dataset is used to monitor the generalisation performance and to avoid overfitting,
while the test dataset provides a final assessment of the trained surrogate amplitude.
The datasets contain approximately $13300$, $1000$, and $1000$ points, respectively.
The evaluation of the hard coefficient at these phase-space points required approximately one million core-hours.
The inputs of the neural network are the ten invariants \mbox{$p_{ij}=p_i\cdot p_j$}, and the output is $H^{(n)}$.
Each input $p_{ij}$ is preprocessed as \mbox{$\hat{p}_{ij}=(\log(p_{ij})-\overline{\log(p_{ij})})/\sigma_{\log(p_{ij})}$},
where $\overline{\log(p_{ij})}$ and $\sigma_{\log(p_{ij})}$ are the mean and standard deviation of $\log(p_{ij})$, respectively.
The MLP uses GELU activation functions~\cite{hendrycks2016gelu} and consists of five hidden layers with 512 neurons each,
for a total of about $10^6$ learnable parameters.
The networks are trained with the Adam optimiser~\cite{kingma2017adammethodstochasticoptimization} and a
cosine-annealing scheduler~\cite{loshchilov2016sgdr} with a maximum learning rate of $2\times10^{-4}$.
We set the maximum number of training iterations to $5\times10^5$ with a batch size of 256~\cite{Breso-Pla:2024pda}.

The accuracy of the surrogate amplitude can be assessed within the ML framework itself.
By the definition of the loss function, replacing the exact amplitude with its surrogate induces an error of at most
$\mathcal{L}\times\int_{\vec{x}\in R}({\as(\muR)}/{(4\pi)})^n|f(\vec{x})|\,\mathrm{d}\vec{x}$ for any phase-space subregion $R$.
The loss value can therefore be regarded as an estimate of the accuracy of the surrogate amplitude.
After the training procedure described above, the test-dataset loss obtained for the exact one-loop amplitude is \mbox{$\mathcal{L}=1.8$\textperthousand}, and those for the amplitudes in LCA are \mbox{$\mathcal{L}=0.30$\textperthousand}
and \mbox{$\mathcal{L}=2.4$\textperthousand}
at the one-loop and two-loop level, respectively.%
\footnote{The lower loss value of the LCA at one-loop level can be understood from the simpler amplitude structure, in particular the absence of the Coulomb singularity that is subleading in colour.
The latter is subtracted in Ref.~\cite{Breso-Pla:2024pda}, which explains the slightly better performance found there.}

\section{Results}
To compare our numerical results with those of Ref.~\cite{Devoto:2024nhl}, we use the setup employed therein.
We consider proton--proton collisions at \mbox{$\sqrt{s}=13.6$\,TeV}.
We use the NNLO NNPDF40 \cite{NNPDF:2021njg} parton distribution functions throughout, with the photon density included, 
and set the CKM matrix diagonal.
The values of the renormalisation and factorisation scales are fixed to \mbox{$\muF=\muR=m_t+m_H/2$}.

\begin{figure}[t]
\includegraphics[width=\textwidth]{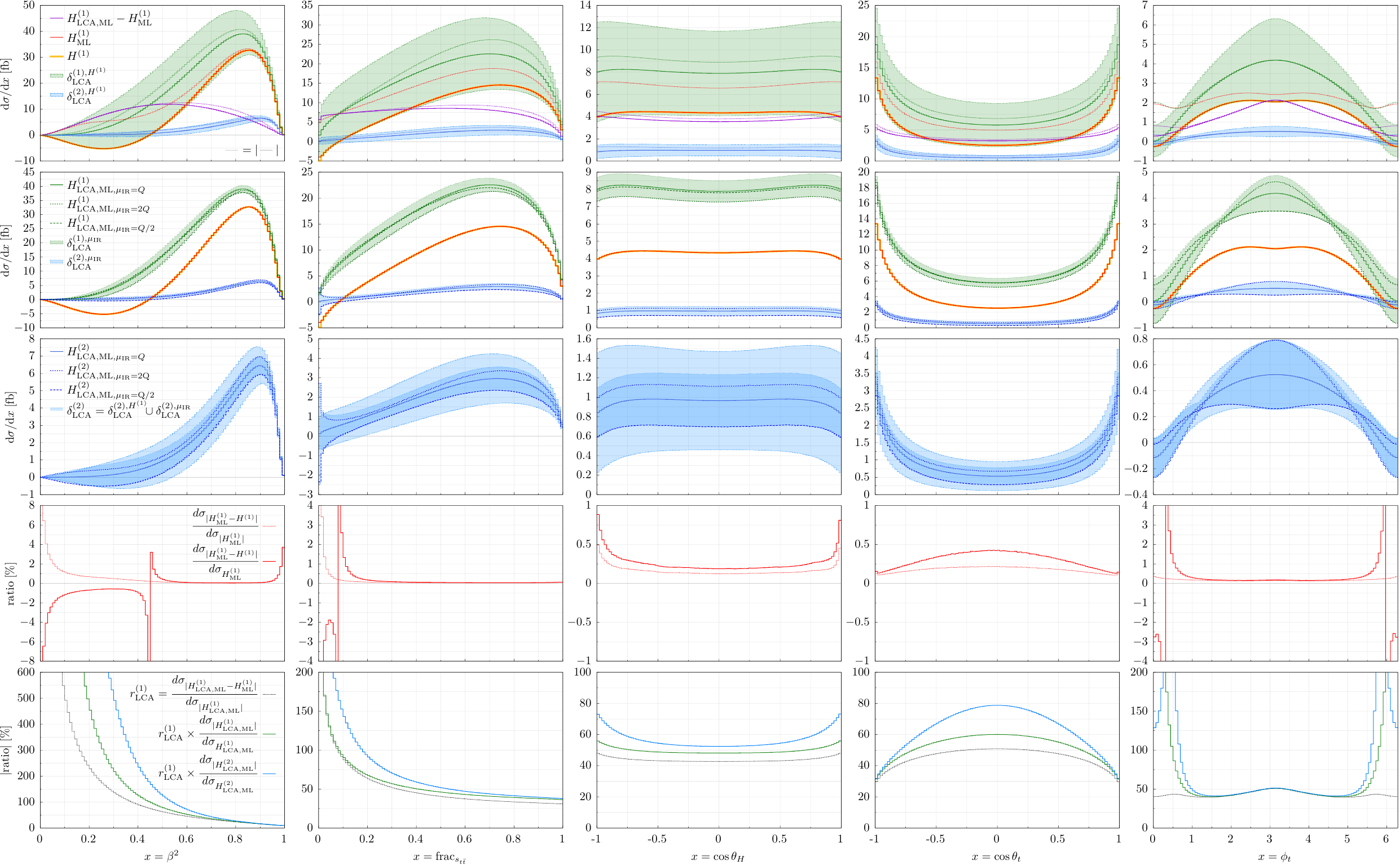}
\caption{\label{fig:results} Distributions in the variables used to parametrise the \mbox{$2\to3$} phase space of \ttH production: $\beta^2$, $\operatorname{frac}_{s_{t\bar t}}$, $\cos \theta_H$, $\cos \theta_t$, and $\phi_t$. A detailed description is given in the main text.}
\end{figure}

The numerical results for the hard-virtual contributions to the cross section are reported in Fig.~\ref{fig:results}
as functions of the five kinematic variables adopted from Ref.~\cite{Agarwal:2024jyq} to
parametrise the phase space, namely two energy fractions,
\betatwo and \fracstt, and three angular variables, \cthetaH, \cthetat, and \phit.
The energy fractions are defined as
\begin{equation}
\label{eq:physvar}
\betatwo = 1 - \frac{(2m_t + m_H)^2}{(p_3+p_4+p_5)^2} \,, \qquad \quad \fracstt = \frac{(p_3+p_4)^2 - 4 m_t^2}{(\sqrt{(p_3+p_4+p_5)^2}-m_H)^2 - 4 m_t^2} \,,
\end{equation}
$\theta_H$ is the polar angle of the Higgs boson in the partonic centre-of-mass frame,
and $\theta_t$ and $\phit$ are the polar and azimuthal angles, respectively, of the
top quark in the $t \bar t$ rest frame.

We start our discussion with the numerical accuracy of the results collected in Fig.~\ref{fig:results}. We first compare the 
hard-virtual contribution based on the exact one-loop amplitude, \dsHone (solid yellow), with that
of the corresponding surrogate amplitude, \dsHoneML (solid red), shown in the first two rows.
As illustrated in the fourth row, we define the deviation of \Hone from \HoneML by taking the modulus of their difference pointwise, i.e.\ before integration,
to suppress cancellations between over- and underestimates of the surrogate amplitude
at different phase-space points.
Since \dsHoneML is not a positive-definite quantity, its relative error (solid red) exhibits spikes at zero transitions
and thus becomes unsuitable as a measure of the approximation quality in these regions.
Taking the modulus pointwise in the normalisation, \dsabsHoneML,
provides instead a reliable estimate of the quality of the surrogate amplitude (dotted red).

We find that \dsHoneML reproduces \dsHone at the level of a few permille in the entire phase space, except for the suppressed
regions $\betatwo\gtrsim0$ and $\fracstt\gtrsim0$, where the deviation reaches several per cent.
Overall, we regard this agreement as completely satisfactory.
Inclusively, we find a deviation of $\sim1.7$\textperthousand{},
in line with the value of the test-dataset loss of $\sim1.8$\textperthousand.
We therefore expect the test-dataset losses of the LCA amplitudes to be good proxies of the respective accuracies as well.

We now turn to the uncertainty of the LCA. On parametric grounds, we expect its accuracy to be ${\cal O}(1/N_c)$. However, cancellations between positive and negative contributions may lead to larger effects.
At the inclusive level, we indeed find $\sHoneLCAML=16.151(5)$\,fb against $\sHone=8.716$\,fb.
Our estimate of the accuracy of the two-loop LCA result is based on the LCA performance at one-loop level, where the exact amplitude is available, and on the ambiguity associated with the choice of the scale \muIR at which the approximation is performed.

The relevant ingredients for the first component are shown in the first row.
As above, we integrate the modulus of the pointwise difference of
\HoneLCAML and \HoneML (dotted violet) and assign it symmetrically as the
absolute error of \dsHoneLCAML, \deltaoneHoneLCA (green band).
For reference, the corresponding deviation defined at the integrated level is shown as well (solid violet),
which corresponds to the difference of
\dsHoneLCAML (solid green) and \dsHoneML (solid red).
Their integrated moduli are also depicted,
\dsabsHoneLCAML (dotted green) and \dsabsHoneML (dotted red),
respectively, to illustrate the extent to which cancellations between positive and negative contributions
occur inside each bin. We observe cancellations across
the full ranges, typically less pronounced for the LCA and most prominent around \mbox{$\phit\sim0$}.

In the fifth row, our measure for the LCA quality at one-loop level,
\mbox{$\roneLCA={\dsabsHoneLCAMLminusHoneML}/{\dsabsHoneLCAML}$}, is shown (dotted black).
It is fairly stable at around \mbox{$40-50\%$} across the angular variables, but exhibits the expected sharp increase
due to the subleading-colour Coulomb singularity as \mbox{$\betatwo\to0$} and, to a lesser extent, as \mbox{$\fracstt\to0$}.
Since the corresponding phase-space region is suppressed, the overall impact
on more inclusive quantities is not dramatic.
Given that the LCA at one-loop and two-loop level is based on an expansion in the same parameter,
we take \roneLCA as an estimate of the LCA uncertainty at two-loop level as well.
To retrieve the standard meaning as relative errors of \dsHnLCAML,
we reweight \roneLCA with the ratio \mbox{${\dsabsHnLCAML}/{\dsHnLCAML}$},
for \mbox{$n=1$} (solid green) and \mbox{$n=2$} (solid blue), respectively.
Back in the first row, we show \dsHtwoLCAML (solid blue)
with the corresponding absolute error, \deltatwoHoneLCA (blue band),
accompanied by \dsabsHtwoLCAML (dotted blue).

The second component of the error estimate reflects the arbitrariness in the choice of the scale \muIR, at which the genuine virtual amplitude is computed, and is illustrated in the second row.
For an exact calculation of \Hn, this choice is irrelevant.
The dependence is exactly compensated for by the running back to the reference scale $Q$.
In an approximation, however, this is no longer the case.
Varying \muIR can provide some insight into the approximation uncertainties.
We compute \dsHoneLCAML at \mbox{$\muIR=Q/2,\;2Q$} (dashed green),
and take the envelope of the symmetrised deviations from \dsHoneLCAML at the nominal scale \mbox{$\muIR=Q$}
as the resulting error estimate, \deltaonemuIRLCA (green band).
We note that this error estimate does not cover the difference with the exact result.
The corresponding results for \dsHtwoLCAML at \mbox{$\muIR=Q/2,\;2Q$} (dashed blue)
and the analogously computed error estimate, \deltatwomuIRLCA (blue band), are reported as well.

The final error estimate assigned to the two-loop contribution \dsHtwoLCAML, \deltatwoLCA, is shown
in the third row, as the envelope of the previously derived components
\deltatwoHoneLCA and \deltatwomuIRLCA.
We find that \deltatwomuIRLCA is typically about a factor of two smaller
than \deltatwoHoneLCA, also at the inclusive level, but comparably large or even exceeding the latter in the regions
\mbox{$\betatwo\gtrsim0$} and \mbox{$\fracstt\gtrsim0$} as well as for \mbox{$\phit\sim0$} and \mbox{$\phit\sim\pi$}.
The shapes of the \dsHoneLCAML and \dsHtwoLCAML distributions are overall qualitatively similar.

With this uncertainty estimate at hand, we can finally turn to the quantitative impact on the inclusive NNLO cross section.
In Ref.~\cite{Devoto:2024nhl}, the $H^{(2)}$ contribution, obtained by combining a soft-Higgs approximation~(SA) and a massification approach~(MA), was found to be
\begin{equation}
\sHtwoSAMA=2.891(1)\,\mathrm{fb} \pm 10.2\%\,,\nonumber
\end{equation}
with the uncertainty explained therein and the integration error stated in parentheses. Our new result is
\begin{equation}
\sHtwoLCA=1.918(5)\,\mathrm{fb} \pm 56.6\%\,,\nonumber
\end{equation}
where the error in parentheses, dominated by using a surrogate amplitude,
is negligible compared to the LCA uncertainty, determined as explained above. This uncertainty is very large, and calls for the inclusion of subleading-colour contributions.

The SA+MA and LCA results are consistent within their respective uncertainties.
This is remarkable, considering that the calculations are done in completely different ways.
Despite the large uncertainty of the LCA, we point out that the difference between the central values of the SA+MA and LCA predictions, about $1\,{\rm fb}$,
is less than two per mille of the inclusive NNLO cross section presented in Ref.~\cite{Devoto:2024nhl}.

\section{Summary}
In this Letter we have presented the first direct calculation of the two-loop QCD
virtual corrections to $\ttH$ production at the LHC.
We focused on the quark--antiquark channel and worked in the generalised leading-colour approximation.
Performing the calculation on a grid of ${\cal O}(10^4)$ points generated according to the LO cross section,
we were able to construct a surrogate amplitude through ML techniques that reproduces the one-loop virtual correction to better than one percent across almost the entire phase space.
We have applied the same method at two-loop order and presented the corresponding results.
The uncertainties of the LCA are quite large and required a detailed discussion.
We have compared our numerical results for the
two-loop virtual contribution to those obtained with dynamical approximations,
finding consistency within their respective uncertainties.  

This work marks an important methodological advance
in the field of two-loop computations. By combining state-of-the-art numerical
methods for amplitude computations with ML techniques, we have shown that
accurate predictions can be obtained for a cutting-edge $2\to 3$ process like \ttH production. 
Our results open the door to NNLO QCD calculations for a broader range of challenging processes,
including other \mbox{$2\to3$}
and potentially even higher-multiplicity processes.

\vspace{0.5cm}
\noindent {\bf Acknowledgments}\\
\noindent
We would like to thank Simone Devoto, Xin Guan, Gudrun Heinrich, Chiara Savoini, and Simone Zoia for valuable discussions and comments on the manuscript.
This work is supported in part by the Swiss National Science Foundation (SNSF) under contract 200020$\_$219367
and by the UZH Postdoc Grant, grant No.~[FK-25-104],
and made use of infrastructure services provided by the Science IT team of the University of Zurich (www.s3it.uzh.ch).

\bibliography{biblio}

\end{document}